\documentstyle[amssymb,multicol,prl,aps,epsfig,amsmath]{revtex}
\begin{document}

\sloppy

\draft

\title{Probing the Strong Boundary Shape Dependence of the Casimir Force}

\author{Thorsten Emig$^{1}$, Andreas Hanke$^{1}$, 
Ramin Golestanian$^{2,4}$, Mehran Kardar$^{1,3}$}
\address{$^{1}$Physics Department, Massachusetts Institute of Technology,
Cambridge, MA 02139\\
$^{2}$Institute for Advanced Studies in Basic Sciences, Zanjan 45195-159, 
Iran\\
$^{3}$Institute for Theoretical Physics, University of California, 
Santa Barbara, CA 93106\\
$^{4}$Institute for Studies in Theoretical Physics and Mathematics,
P.O. Box 19395-5531, Tehran, Iran}

\date{\today} 
\maketitle

\begin{abstract}
We study the geometry dependence of the Casimir energy for
deformed metal plates by a path integral
quantization of the electromagnetic field. For the first time, we
give a complete analytical result for the deformation induced change
in Casimir energy $\delta{\cal E}$ in an experimentally testable,
nontrivial geometry, consisting of a flat and a corrugated plate. Our
results show an interesting crossover for $\delta {\cal E}$ as a
function of the ratio of the mean plate distance $H$, to the corrugation
length $\lambda$: For $\lambda \ll H$ we find a {\em slower} decay
$\sim H^{-4}$, compared to the $H^{-5}$ behavior predicted by the
commonly used pairwise summation of van der Waals forces, which is
valid only for $\lambda \gg H$.
\end{abstract}


\begin{multicols}{2}\narrowtext

In 1948, Casimir showed that two parallel conducting plates, separated
by a distance $H$, attract each other with a force $F$, proportional to 
the surface area $A$, and given by \cite{Casimir48}
\begin{equation} \label{flat}
\frac{F}{A} \, = \, - \, \frac{\pi^2}{240} \frac{\hbar c}{H^4} \, \, .
\end{equation}
This remarkable prediction of quantum electrodynamics can be
understood as resulting from the modification of the zero point vacuum
fluctuations of the electromagnetic field by the presence of
boundaries.  Since this discovery, the fundamental nature of the
Casimir effect and its implications, e.g., on surface forces
\cite{Israelachvili92}, particle physics \cite{Milton80}, and
cosmology \cite{BK84}, has motivated extensive theoretical work (see
Refs.~\cite{PMG86,MT97,ER91,Milonni94,Kardar99} for reviews).  On the
experimental front, the initial attempt at observing the Casimir
force, by Sparnaay in 1958, was not conclusive due to large
experimental uncertainty \cite{Sparnaay58}. Only recently, there have
been a number of precision measurements of the Casimir force, using a
torsion pendulum \cite{Lamoreaux97}, an atomic force microscope
\cite{MR98}, and a micromachined torsional device \cite{CAKBC2001},
which confirm the theory to a few per cent accuracy. The latter
experiment also demonstrates the possibility for novel actuation
schemes in microelectromechanical systems based on the Casimir force
\cite{SWM95}.

In the more general context of the Lifshitz theory for dielectric
bodies \cite{Lifshitz}, Eq.\,(\ref{flat}) appears in the limit of
perfectly conducting plates, for which the dielectric constant
$\varepsilon$ is infinite.  For finite $\varepsilon =
\varepsilon(\omega)$, this power law for the force is recovered for
large distances $H \gg c/\omega_0$, where $\omega_0$ is the smallest
resonance (absorption) frequency of the dielectric (usually
$c/\omega_0\approx 10 \text{--} 100 \, \mbox{nm}$). 
In this, so-called retarded,
limit, the force is {\em universal\/} in the sense that it only
depends on the electrostatic dielectric constant $\varepsilon_0 =
\varepsilon(0)$, and can be obtained, e.g., by dispersion relation
techniques \cite{Feinberg+70}. The opposite limit of $H \ll c/\omega_0$
gives the unretarded van der Waals force $F/A \sim H^{-3}$, which can
also be obtained by summing the (attractive) intermolecular
interactions due to induced molecular dipole moments. Even though
obtainable from the same microscopic theory, the Casimir and van der
Waals forces are quite different. In particular, the interpretation of
the Casimir force in terms of changes in zero point vacuum
electromagnetic energy suggests it to be a strong function of geometry
\cite{Balian,source}; probing the global shape of the boundary that
confines the vacuum fluctuations.  Indeed, whereas the van der Waals
force between electrically polarizable particles is always attractive,
even the {\em sign\/} of the Casimir force is geometry dependent, and
can be {\em repulsive\/}, e.g., for a thin spherical or cubic
shell \cite{MT97,ER91,Balian}. (Repulsive
Casimir forces are expected also when {\em magnetic} as well as
electric properties are included \cite{Feinberg+70,Boyer+Hushwater}.)
%
%
\unitlength1cm
\vspace*{-.6cm}
\begin{figure}[t]
\begin{picture}(8,4.5)
\put(-0.5,0.7){
\setlength{\epsfysize}{3.5cm}
\epsfbox{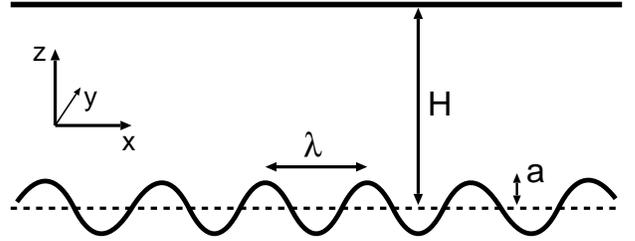}}
\end{picture}
\vspace*{-.6cm}
\caption{Geometry used for calculating the Casimir energy of a
flat plate and a corrugated plate at mean distance $H$.}
\vspace*{-.3cm}
\label{fig1}
\end{figure}
%

\begin{figure}
\includegraphics[width=0.9\linewidth]{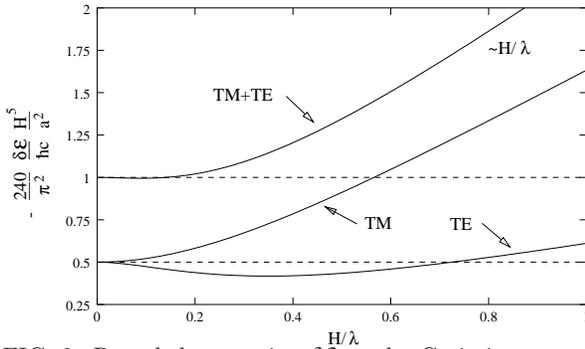}
\caption{Rescaled correction $\delta {\cal E}$ to the Casimir energy
due to the corrugation as given by the terms in square brackets of
Eq.~(\ref{ce}) (upper curve).  The lower curves show the separate
contributions from TM and TE modes.  The rescaling of $\delta {\cal
E}$ is chosen such that the result from a pairwise summation of van
der Waals forces is a constant (dashed lines).}
\label{fig2}
\end{figure}

It is highly desirable to demonstrate the strong shape dependence of
the Casimir interaction in a set-up that clearly demonstrates its
distinction from the usual pairwise additive 
interactions \cite{Kardar}. Since
measurement of the repulsive Casimir interaction for a conducting sphere is
experimentally difficult, the most promising route is to via
modifications of the parallel plate geometry.  In searching for
nontrivial boundary dependences, Roy and Mohideen \cite{exp1} examined
the force between a sphere, and a sinusoidally corrugated plate with
amplitude $a\approx 60 \, \mbox{nm}$ and wavelength 
$\lambda\approx 1.1 \, \mu\mbox{m}$. Over the range of separations 
$H\approx 0.1 - 0.9 \, \mu\mbox{m}$, the observed
force showed clear deviations from the dependence expected on the
basis of decomposing the Casimir force to a sum of pairwise
contributions (in effect, an average over the variations in
separations).  This experimental result  motivated our
calculation of the exact Casimir force in the geometry depicted in
Fig.\,\ref{fig1}, without the assumption of pairwise additivity.  
Our analytic results [see Eqs. (4) - (6) and Fig.\,\ref{fig2}] 
hold to second order in $a$, and
show that for fixed $H$ the corrections due to corrugation strongly
depend on $\lambda$.  In fact, for $H/\lambda\gg 1$ the correction is
by a factor of $H/\lambda$ {\em larger} than in the opposite limit of
$H/\lambda\ll 1$ where the assumption of pairwise additivity is
asymptotically correct.  However, the experiments of
Ref.\,\cite{exp1} are performed in the range of $H/\lambda\approx 0.1
- 0.8$ where the corrections to pairwise additivity are in fact not
significant enough to account for the observed deviations.  This
bolsters the conclusion in Ref.\,\cite{exp2} that these deviations are
in fact due to a lateral force that tends to preferentially position
the spherical AFM tip on top of local maxima of the modulated surface
(leading to a smaller separation and stronger force).  We thus propose
that the shape dependence of the Casimir force can in fact be probed in
this set-up by going to modulations of shorter wavelength; a hard but
achievable goal.  

The use of a spherical tip, of large radius $R$, in
experiments \cite{Lamoreaux97,MR98,CAKBC2001} causes some differences
from the flat plate geometry used in our calculations.  First, the
positioning of the tip relative to the modulations is important when
$H$ and $\lambda$ are comparable, but becomes insignificant in the
proposed limit of $\lambda \ll H,R$.  
Secondly, as long as $R\gg H, \lambda$
the curvature of the tip does not lead to nontrivial corrections, and
the force can be related to the energy per surface area $\cal E$ in
Eq.\,(5) by the proximity force rule $F=2\pi R\cal E$ \cite{D34}. 
These formulea thus provide a specific recipe for evaluating the
nontrivial shape
dependences of the Casimir force in the 
experimental 
set-up.

Before turning to the geometry of Fig.~\ref{fig1},  consider the
more general case of two perfectly conducting plates of mean
separation $H$, which are infinitely extended along the $x$-$y$
plane. Now assume that one of the plates is deformed in a way that is
translationally invariant along the $y$ axis and has no overhangs.
Its profile can then be described by the height function $h(x)$,
with $\int dx\, h(x)=0$. For example, to describe the
geometry in the experiment of Ref.\cite{exp1}, we choose
$h(x)=a\cos(2\pi x/\lambda)$
as in Fig.~\ref{fig1}. The Casimir energy associated with general
$h(x)$ at zero temperature corresponds to the difference of the
ground state energies of the quantized electromagnetic (EM) field with
and without plates, respectively. To obtain this energy, we employ the
path integral quantization method, which can be applied to the EM
gauge field by introducing a suitable gauge fixing procedure
\cite{PS-book}. However, in the present translationally invariant
geometry we can develop a simpler quantization scheme, by a similar
reasoning as used in the context of waveguides with constant
cross-sectional shape (here along the $y$ axis) \cite{Jackson}. For
an arbitrary EM field between the plates,  transverse components
of the $E$ and $B$ fields are completely determined by their axial
components $E_y$ and $B_y$. Therefore, any EM field can be described
by a superposition of two 
independent scalar fields $\Phi_{\rm TM}\equiv E_y$ (transverse
magnetic waves, $B_y= 0$) and $\Phi_{\rm TE}\equiv B_y$
(transverse electric waves, $E_y= 0$). The scalar fields both
fulfill the usual wave equation, but differ in their boundary
conditions on the plates S, as
$\Phi_{\rm TM}|_{S} = 0$, while
$\partial_n \Phi_{\text{TE}}|_{S}=0$,
where $\partial_n$ denotes the normal derivative. 

Both scalar fields can now be quantized  by considering the
Euclidean action
$S[\Phi]=\frac{1}{2} \int d^4 X \,(\nabla \Phi)^2$,
corresponding to the wave equation after a Wick rotation to the
imaginary time $X^0=ict$. In the 4D Euclidean space, the plates are
parametrized by $X_1({\bf r})=[{\bf r},h(x)]$ 
and $X_2({\bf r})=[{\bf r},H]$, 
with ${\bf r}=(ict,x,y)$. 
We implement the boundary conditions on $S$ using delta functions
\cite{Kardar,KN2001}, leading to the partition function
\begin{equation}
{\cal Z}=\frac{1}{{\cal Z}_0}\int {\cal D}\Phi \prod_{j=1}^2 C[\Phi(X_j)]
\exp(-S[\Phi]/\hbar),
\end{equation}
with the boundary condition enforcing functionals
$C[\Phi(X_j)]=\prod_{\bf r} \delta(\Phi(X_j({\bf r}))$ for $\Phi_{\rm
TM}$, and $C[\Phi(X_j)]=\prod_{\bf r} \delta(\partial_n \Phi(X_j({\bf
r}))$ for $\Phi_{\rm TE}$, and the partition function ${\cal Z}_0$ of
the space without plates. The Casimir energy per surface area $A$ is
then given by ${\cal E}=-\hbar c \ln{\cal Z}/AL$ where $L$ is the
overall Euclidean length in time direction. Implementing the delta
functions by integrals over  auxiliary fields and integrating out
$\Phi$, we obtain an $h(x)$ dependent kernel $M$ for the Gaussian
action of the auxiliary fields. Expanding ${\cal Z}=({\rm det}
M)^{-1/2}$ to second order in $h(x)$, we get for the deformation
dependent part of ${\cal Z}$,
\begin{equation}
\ln {\cal Z}_h =
\frac{1}{2} \int_{\bf r} \int_{{\bf r}'}\!
K({\bf r}- {\bf r}') 
h(x) h(x').
\end{equation}
The kernel is the sum of contributions from the two wave types, i.e.,
$K({\bf r})=K_{\rm TM}(|{\bf r}|)+K_{\rm TE}(|r_0|,|{\bf r}_\parallel|)$,
with ${\bf r}=(r_0,{\bf r}_\parallel)$, and has been calculated explicitly.

\begin{figure}
\includegraphics[width=0.9\linewidth]{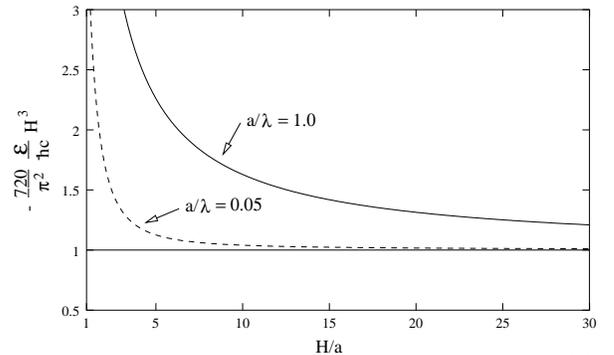}
\caption{Rescaled Casimir energy as given by Eq.~(\ref{ce}) for two
fixed values of $a/\lambda$. The rescaling is chosen such that the
Casimir energy of two flat plates becomes one (horizontal line).  The
lower curve with $a/\lambda=0.05$ corresponds to the parameters used in
the experiment of Ref.~\protect\cite{exp1}, where $H/a$ varies between
approximately 3 and 17.}
\label{fig3}
\end{figure}

For the specific deformation of the plates corresponding to harmonic
corrugation of amplitude $a$ and wavelength $\lambda$ defined above, 
the calculation of $\ln {\cal Z}_h$ 
reduces to Fourier transforming the kernel $K({\bf r})$. The
corresponding integrals can be performed for $\lambda>0$ by closing the
integration contour via a semi-circle at infinity in the upper
half of the complex plane. The resulting sum of an infinite series
of residues can be expressed in terms of the
polylogarithm function ${\rm Li}_n(z)\equiv\sum_{\nu=1}^\infty
z^\nu/\nu^n$, leading to
\begin{equation}
\label{ce}
{\cal E}=-\frac{\hbar c}{H^3}\left\{
\frac{\pi^2}{720}+\frac{a^2}{H^2}\left[
G_{\rm TM}\left(\frac{H}{\lambda}\right)+
G_{\rm TE}\left(\frac{H}{\lambda}\right)\!
\right]\!\right\}.
\end{equation}
The contributions from TM and TE modes are:
\end{multicols}
\widetext
\begin{eqnarray}
\label{tm}
G_{\rm TM}(s)&=&\frac{\pi^3s}{480}-\frac{\pi^2 s^4}{30} \ln(1-u)
+ \frac{\pi}{1920 s} {\rm Li}_2(1-u)
+ \frac{\pi s^3}{24} {\rm Li}_2(u) 
+ \frac{s^2}{24} {\rm Li}_3(u) + \frac{s}{32\pi} {\rm Li}_4(u)\nonumber\\ 
&& + \frac{1}{64\pi^2}
{\rm Li}_5(u)+\frac{1}{256\pi^3 s} \left( {\rm Li}_6(u)-
\frac{\pi^6}{945}\right),\label{gtm}\\[5mm]
\label{te}
G_{\rm TE}(s)&=&\frac{\pi^3 s}{1440}-\frac{\pi^2 s^4}{30} \ln(1-u)
+\frac{\pi}{1920 s} {\rm Li}_2(1-u)
-\frac{\pi s}{48}\left(1+2s^2\right){\rm Li}_2(u)+\left(\frac{s^2}{48}-
\frac{1}{64}\right){\rm Li}_3(u)+\nonumber\\
&& + \frac{5s}{64\pi}{\rm Li}_4(u)+\frac{7}{128\pi^2} {\rm Li}_5(u)
+\frac{1}{256\pi^3s}\left(\frac{7}{2}{\rm Li}_6(u)-\pi^2 {\rm Li}_4(u)
+\frac{\pi^6}{135}\right),
\label{gte}
\end{eqnarray}
\vspace*{-0.5cm}
\begin{multicols}{2}
\narrowtext
\noindent
with $u\equiv \exp(-4\pi s)$. Figure~\ref{fig2} displays separately
the contributions from $G_{\rm TM}$ and $G_{\rm TE}$ to the corrugation 
induced correction $\delta {\cal E}$ to the Casimir energy.  
While $G_{\rm TM}(H/\lambda)$ is a monotonically increasing function
of $H/\lambda$, $G_{\rm TE}(H/\lambda)$ displays a minimum 
for $H/\lambda \approx 0.3$.  
The net Casimir energy ${\cal E}$ is shown in Fig.~\ref{fig3} for two
representative values of $a/\lambda$, including the parameters used in
the experiment of Ref.~\cite{exp1}. Note that the corrugation induced
correction leads to a larger energy ${\cal E}$, and hence the
corresponding force $F=2\pi R {\cal E}$ is {\em enhanced}.
Examining the the limiting behaviors of Eq.~(\ref{ce}) is instructive. 
In the limit $\lambda \gg H$, the functions $G_{\rm TM}$ and $G_{\rm TE}$ 
approach constant values,
and the Casimir energy takes the $\lambda$-independent form
\begin{equation}
{\cal E}=-\frac{\hbar c}{H^3}\frac{\pi^2}{720}
\left(1+3\frac{a^2}{H^2} \right). 
\end{equation}
Note that only in this case both wave types provide the same
contribution to the total energy, see Fig.~\ref{fig2}. 
In the opposite limit of $\lambda \ll H$, as suggested by the first terms in
Eqs.~(\ref{gtm}), (\ref{gte}), both $G_{\rm TM}$ and $G_{\rm TE}$ grow
linearly in $H/\lambda$. 
Therefore, in this limit the correction to the Casimir energy
decays {\em slower}, according to a new power law in $H$,
\begin{equation}
\label{large-H}
{\cal E}=-\frac{\hbar c}{H^3}\frac{\pi^2}{720}
\left(1+2\pi\frac{a^2}{\lambda H}\right),
\end{equation}
with an amplitude proportional to $1/\lambda$.  Analyzing the
correction $\delta {\cal E}$ in the limit $a,\lambda \ll H$ for {\em
arbitrary} values of $a/\lambda$, we find that the factor multiplying
$a/H$ in Eq.~(\ref{large-H}) saturates for $\lambda \ll a$ at a number
of order unity.  This result can be justified by noting that the most
relevant contributions to the force come from modes of wavelength of
order $H$. The corrugation also affects modes of wavelength of order
$\lambda$, but these modes contribute to the single plate energy only.
Thus, in the extreme limit $\lambda \ll a$, one has a clear separation
of the length scales $H$ and $\lambda$, and the modes ``see" flat
plates at an effective separation $H-a$, leading to a correction of the
order $a/H$ after expansion in $a$.

The above behavior of the correction $\delta{\cal E}$ for
small and large $H/\lambda$ clarifies the limits of validity of previous
results in the literature. The upper dashed line in
Fig.~\ref{fig2} corresponds to a widely used approach
\cite{Israelachvili92,vdW} in which the interaction is obtained from
a pairwise summation of `van der Waals type' two body forces. It is
evident that this approximation is accurate only for $H/\lambda \to
0$, which in this limit is equivalent to the Derjaguin method to any
order in the amplitude $a$ \cite{D34}. Already for $H/\lambda$ of
order unity, the additive van der Waals type approximation breaks down.
The opposite limit, $H/\lambda \to \infty$, corroborates the result
reported in Ref.~\cite{KNS87}, which is larger than the former by a
factor of $H/\lambda \gg 1$.  However, in experiments with lateral
distortions $\lambda$ of the order of $H$, none of the above limiting
cases is realized, which makes the present, more complete analysis
necessary.

Moreover, for purposes of experimental comparison, corrections due
to finite conductivity of the plates, surface roughness, and finite
temperature should be taken into account. These corrections introduce
additional length scales into the problem, which are in turn the
plasma wavelength $\lambda_p$ of the plates (e.g., $\lambda_p \approx
100 \, \mbox{nm}$ for aluminium \cite{MR98,exp1}), the transverse
correlation length $\xi$ of the roughness (usually $\xi \approx 300 \,
\mbox{nm}$ \cite{CAKBC2001}), and the thermal wavelength
$\lambda_T=\hbar c / k_B T$ ($\approx 1 \, \mu\mbox{m}$ at $300^\circ \,
\mbox{K}$).  The plasma and thermal wavelengths provide lower and
upper bounds for $H$, respectively, such that our results for perfectly
conducting plates at zero temperature are valid for $\lambda, H \gg
\lambda_p$, and $H \ll \lambda_T$ \cite{Roughness}.

Finally, we note that in the set-up of Fig.~\ref{fig1} nontrivial shape
dependencies appear as corrections to a larger Casimir force.  For the
purpose of experimental tests, it is much more desirable to devise
set-ups which directly probe differences, without the need for
subtracting a larger baseline force.  For example, in an
atomic force experiment, simultaneous scanning of a flat and corrugated
substrate would be desirable; while in the torsion
pendulum experiment, one can imagine suspending a spherical lens
equidistantly from two plates, one of which is corrugated.  Another potential
experiment along these lines is to measure the lateral force between
two plates with sinusoidal corrugations of the same wavelength
$\lambda$, which are shifted relative to each other by a distance
$\delta$.  We find a lateral force $F_\|=\hbar c \frac{a^2}{\lambda
H^5} \sin(2 \pi \delta /\lambda) \, g(H/\lambda)$.  This force tends
to change the position of the plates such that a maximum is opposite
to a minimum, corresponding to $\delta = \lambda/2$.  The universal
function $g(H/\lambda)$ tends to a finite value for $H/\lambda \to 0$,
but vanishes exponentially for large $H/\lambda$.  
Similar results hold for the alternating component of the standard
Casimir force as a function of the phase shift between the plates.

We thank J.-P.~Bouchaud, S.~Dietrich, S.~G.~Johnson, M.~L.~Povinelli,
and S.~Scheidl for useful discussions. This work was supported by the
Deutsche Forschungsgemeinschaft under grants No. EM70/1-3 (T.E.),
$\mbox{HA3030/1-2}$ (A.H.), and the National Science Foundation
through grants No. DMR-01-18213, and PHY99-07949 (M.K.).

\end{multicols}


\begin{references}
\bibitem{Roughness}
The importance of stochastic surface roughness can also be deduced from 
our calculations.  The relative corrections $\delta {\cal E}/{\cal
E}$ to the Casimir energy due to  roughness of
amplitude $a$ and  transverse correlation length $\xi$ should be
of the form $a^2/H^2$ for $\xi \gg H$, and  $a^2/(\xi H)$
for $\xi \ll H$.  While the latter behavior is in accordance with
Ref.~\cite{NSC}, we note that the experimental case corresponds to 
neither extreme, making a more complete analysis necessary.
\end{references}
\end{document}